\def\npb{Nucl. Phys. B}
\def\plb{Phys. Lett. B}
\def\prl{Phys. Rev. Lett. }

\def\be{\begin{equation}}
\def\ee{\end{equation}}
\def\ba{\begin{eqnarray}}
\def\ea{\end{eqnarray}}

\def\mathrm{\rm}
\documentstyle[12pt]{article}
\begin{document}

\title{{\bf Four Fermions Productions at a $\gamma\gamma$ Collider}}

\author{
{\bf Mauro Moretti,}
\\ \it (e-mail: moretti@vaxfe.fe.infn.it)
}
\date{}
\maketitle

\begin{abstract}
Using the recently proposed ALPHA algorithm (and the resulting code)
I compute the rate (at tree level) for the process 
$\gamma\gamma\rightarrow\bar\nu_e e^- u \bar d$.
The bulk of the contribution is due to W pair production and
decay. However a non negligible ($\sim10 \%$) contribution 
comes from other channels, mainly the production
and decay of a W and a collinear charged fermion.
Requiring that the reconstructed
invariant $u \bar d$ mass lies in the intervals $M_W\pm 5 $ GeV
and $M_W\pm 20 $ GeV one obtains a rate 
which is lower, by 25 \% and 4 \% respectively,
than the rate obtained in the $narrow$ $width$
approximation, thus demonstrating the relevance of
the finite W width.
\end{abstract}

%\hspace{12cm} SHEP 95-22
%\vspace{15cm}
\section{Introduction}

Future high energy $e^+ e^-$ colliders \cite{nlc}
are likely to make available the possibility
to operate also in the $e\gamma$
and $\gamma\gamma$ mode \cite{gcol}.
This last mode will allow the experimental study
of the bosonic sector of the electroweak lagrangian
\cite{selfbos},
triple and quartic gauge boson couplings as well
as the coupling of gauge bosons with the
higgs particle if it is light enough to be produced.

The most abundant final state to be studied will be 
W pair production. This process is allowed at the tree level and,
since the production rate is dominated by $t$ channel
virtual W exchange, the cross section becomes nearly
 constant  for a center of mass
energy above $400$ GeV. With the aimed luminosity in the range of
 $10 \div 20$ inverse femtobarns per year
about one millions of W pairs per year are expected.

Because of the high statistic and of the relatively clean
environment provided by a leptonic collider, a good 
accuracy in the theoretical prediction will be necessary.

Since the W boson is shortlived,
the experimental signature for W pair production
is via its decay products, 
mostly four fermions in the final state.
Therefore one has to compute the 
rate for the process $\gamma\gamma\rightarrow 4 \ fermions$.

In this paper I
compute, at tree level, the rate for 
the process
\be
\gamma \gamma \rightarrow \bar \nu_e e^-
u \bar d
\label{pro}
\ee
discussing in some detail the comparison
with the {\it narrow width} approximation and
with the calculation which 
includes only the contribution
of the subset of doubly resonant diagrams.
I carry on the comparison both at the level of total cross section
and of differential distributions.

\section{The computation}
The  amplitude for the process (\ref{pro})
 is computed using a new technique which,
in collaboration with {\it F. Caravaglios} \cite{alpha},
I have recently developed.
Exploiting the relation between the one-particle irreducible Green 
Functions generator $\Gamma$ and the connected Green Functions
generator $G$ we have proposed a simple numerical algorithm to compute
tree level scattering amplitudes.
We have then implemented the algorithm in a FORTRAN code ALPHA
which presently uses the standard electroweak lagrangian (QCD is
not included yet) and can compute any scattering
amplitude in this framework,
in a fully automatic way.  
We have used the ALPHA code to obtain the rates for
the processes $e^+ e^-$ {\it into four fermions}
which are of relevance for the LEP200 experiments and
our results are in complete agreement, within the statistical error
of the Montecarlo integration
(typically less than one per mille), with
the results obtained using more conventional methods
\cite{lep200}.
This provides the most stringent test of our algorithm
and of our code:
it is important to stress 
here  that the
calculation of the matrix element 
of any of these processes 
is entirely automatic;
to study a different process we had to change only an input
file specifying the type and number of particles
involved.

The input values,
which will be used in
the present paper,
 for particles masses, widths and
for the electroweak coupling constants are reported in table \ref {ew}.
The running width scheme is used, namely the W propagator
$\pi_W$ is taken
as
\be
\pi_W^{\mu\nu} = \frac { - i (g^{\mu\nu} - p^\mu p^\nu 
/M_W^2) }
{p^2 - M_W^2 + i \Gamma_w p^2 \theta (p^2)  /M_W}
\label{rw}
\ee
where $p$ is the W four momentum, $M_W$ and $\Gamma_W$ are
the W mass and width respectively,
and $\theta(p^2)$ is the Riemann $\theta$ function:
it is equal to one for positive $p^2$
and zero otherwise.

\begin{table}
\begin{center}
\begin{tabular}{|c|c|c|c|}
\hline\hline
$M_W=80.23$ GeV & 
$\Gamma_W = 2.03367 $ & 
$\alpha_{QED}=1./128.07$ & 
$ \sin^2 \theta_W=0.23103$ \\ 
\hline 
$m_u=5$  MeV & 
$m_d=10$  MeV & 
$m_e=0.51$  MeV & 
   \\
\hline  \hline
\end{tabular}
\end{center}
\caption {
Input parameters for the electroweak lagrangian.
$m_e$,  $m_u$, $m_d$  are the electron $u$ and $d$ quarks
masses respectively, $M_W$ and $\Gamma_W$
the W mass and width, $\theta_W$ is the Weinberg angle
and $\alpha_{QED}$ is the electromagnetic coupling constant.
Tree level relationships among the parameters of
the standard model electroweak lagrangian are assumed.
}
\label{ew}
\end{table}

The bulk of the contribution to the rate for (\ref{pro})
is obtained when the process is mediated by two almost on shell W
which decay into two pairs of fermions or when one of the charged
fermions
is collinear to one of the incoming photons
and an on shell W is emitted.
 Therefore, to perform the numerical integration over
the phase space variables, one
 needs to increase  the sampling in these two regions.
To this purpose  I have used the
package VEGAS \cite{vega}
and all the reported results are obtained with at least
twenty VEGAS estimates of the integral with a $\chi^2$
smaller than two.

To assess the performances of the ALPHA code as 
an event generator for $\gamma\gamma$ processes the
CPU time required for the evaluation of the scattering matrix element
of several final states is reported in table \ref{cpu}.

\begin{table}
\begin{center}
\begin{tabular}{|c|c|}
\hline\hline
Process & CPU time in seconds per $10^6$ events \\
\hline \hline
$\gamma\gamma \rightarrow \nu_e e^- u \bar d $ (resonant) 
& 2208
\\
\hline
$\gamma\gamma \rightarrow \bar \nu_e e^- u \bar d $ (all)
& 4960
\\
\hline
$\gamma\gamma \rightarrow  \gamma \bar \nu_e e^- u \bar d $
& 12800
\\ \hline
$\gamma\gamma \rightarrow \bar 
\nu_e e^- u \bar d e^- e^+$ 
& 51600 \\ \hline 
$\gamma\gamma \rightarrow \bar \nu_e e^- \nu_e e^+ 
 e^- e^+$ & 84400 \\ \hline 
  \hline
\end{tabular}
\end{center}
\caption {
CPU time required by the ALPHA code for the evaluation of some
final states from $\gamma\gamma$ fusion. Times are in second
per $10^6$ events. These performances are obtained with 
a DIGITAL machine
ALPHA 3000/600 with 64M of memory. Double precision is used.
All fermions are massive.
}
\label{cpu}
\end{table}

\section{The Results}

One of the most important (and abundant) process to be observed
at the proposed $\gamma\gamma$ collider will
be W pair production
\be
\gamma\gamma \rightarrow W^+ W^-.
\label{ww}
\ee
Because of the finite (and `short') lifetime of the W,
at the collider one will observe 
the W's decay products, mostly four final fermions.
As a first order approximation (the so called {\it narrow width}
approximation)
one computes the rate for the process
(\ref{ww}) and multiplies it for
the branching ratios appropriate to the given four fermions
final state.

My purpose here is to confront this approximation
with the results for the full computation of the process
(\ref{pro}) at tree level.
There are two main differencies:
in addition to the diagrams accounting for W pair
production and decay,
a lot more diagrams contribute to the same final state
and, because of the finite W width, the bulk of the cross section
occurs for
a reconstructed W invariant mass
spread over a few W widths,
making the definition of a `W' via invariant
mass cuts more delicate
and affecting in a sizable way the cross section.

At a realistic $\gamma\gamma$ collider the photon energy
spectrum will not be nearly monochromatic (as the one of the
parent $e^+e^-$) and, to account for the experimental features,
one should fold the photon spectra with the cross section
for (\ref{pro}). However,
for the present  purposes, a discussion of the main features
at a fixed center of mass energy is sufficient.

In Fig.~1 the differential
cross section, at a center of mass energy of 500 GeV,
for the process (\ref{pro}) is
plotted as a function of $\mu={\mathrm max} (\mu_1,\mu_2)$
where $\mu_1=|[(p_e+p_\nu)^2]^{1/2} - M_W|$
$\mu_2=|[(p_u+p_d)^2]^{1/2} - M_W|$
and $p_e$, $p_\nu$, $p_u$, $p_d$ 
are the electron, neutrino u and d quarks momenta
respectively. Some numerical values for the same quantities
are reported in table \ref{tggcut}. In table \ref{tggcut1}
the same quantities are given for center of mass energies of
300 and 1000 GeV.
I give the results for the full set of diagrams
and those obtained using only resonant contributions, 
namely those accounting for W pair production and decay.

\begin{table}
\begin{center}
\begin{tabular}{|c|c|c|c|c|c|}
\hline\hline
   Diagrams  & Angular Cut 
& $\sigma (5) \;(pb)$ & $\sigma (10) \;(pb)$ & $\sigma (18) \;(pb)$
& $\sigma (250) \;(pb)$ 
  \\
\hline\hline
 $all$  &  $ |\cos\theta_f| < 1 $ 
& 2.529(3) & 2.932(3) & 3.153(3) & 3.825(4)
\\
\hline $resonant$  &  $ |\cos\theta_f| < 1 $ 
& 2.508(2) & 2.886(2) & 3.068(2) & 3.374(2)
\\
\hline $all$ $(fudge)$  &  $ |\cos\theta_f| < 1 $ 
& 2.524(3) & 2.926(3) & 3.146(3) & 3.797(4)
\\
\hline $all$  &  $ |\cos\theta_f| < 0.98 $ 
& 1.748(2) & 2.012(2) & 2.138(2) & 2.290(2) 
\\
\hline $resonant$  &  $ |\cos\theta_f| < 0.98 $ 
&1.753(2) & 2.016(2) & 2.144(2) & 2.378(2)
\\
\hline $all$ $(fudge)$ &  $ |\cos\theta_f| < 0.98 $ 
& 1.752(2) & 2.014(2) & 2.142(2) & 2.294(2) 
\\
\hline $all$  &  $ |\cos\theta_f| < 0.92 $ 
& 0.7575(9)
& 0.871(2) & 0.926(1) & 0.993(1)
\\
\hline $resonant$  &  $ |\cos\theta_f| < 0.92 $ 
& 0.759(1) & 0.874(1) & 0.930(1) & 1.053(1)
\\
\hline $all$ $(fudge)$  &  $ |\cos\theta_f| < 0.92 $ 
& 0.760(2) &0.867(2) & 0.929(2)
&0.996(2)
\\
\hline \hline
\end{tabular}
\end{center}
\caption {
Rates for $ \gamma \gamma \rightarrow \nu_e e^-
u \bar d
$ for a center of mass energy of 500 GeV.
$\sigma(a)$ is the cross section with the invariant mass
cut $\mu < a$ (GeV).
The variable $\mu$ is defined as the maximum value
of
 $\mu_1=|[(p_e+p_\nu)^2]^{1/2} - M_W|$ and
$\mu_2=|[(p_u+p_d)^2]^{1/2} - M_W|$
and $p_e$, $p_\nu$, $p_u$, $p_d$ 
are the electron, neutrino u and d quarks momenta
respectively. The angle $\theta_f$ is the minimal angle
of charged fermions with respect to the beam direction.
Entries labelled $all$ refer to the full computation
(all Feynman Graphs included),
those labelled $resonant$ to the resonant one 
(only doubly resonant Feynman Graphs included) and those
labelled $fudge$ to the full computation using the
$fudge$ scheme (see eq.~(5) ) with the running width.
The rate in the {\it narrow width }approximation (3) 
is $3.29 \ pb$.
}
\label{tggcut}
\end{table}

\begin{table}
\begin{center}
\begin{tabular}{|c|c|c|c|c|c|c|}
\hline\hline
Energy &   Diagrams  & Angular Cut &  
 $\sigma (5) \;(pb)$ & $\sigma (10) \;(pb)$ & $\sigma (18) \;(pb)$
& $\sigma (250) \;(pb)$ 
  \\
\hline\hline
300 & $all$  &  $ |\cos\theta_f| < 1 $ 
& 2.291(3) & 2.648(3) & 2.838(4) & 3.329(9)
\\
\hline
300 & $resonant$  &  $ |\cos\theta_f| < 1. $ 
& 2.274(2) & 2.616(2) & 2.780(2) & 2.982(2)
\\
\hline
300 & $all$  &  $ |\cos\theta_f| < 0.98 $ 
& 1.976(2) & 2.273(2) & 2.416(3) & 2.574(3)
\\
\hline
300 & $resonant$  &  $ |\cos\theta_f| < 0.98 $ 
& 1.976(2) & 2.274(2) & 2.417(2) & 2.596(2)
\\
\hline
300 & $all$  &  $ |\cos\theta_f| < 0.92 $ 
& 1.330(2) & 1.529(2) & 1.623(3) & 1.718(3)
\\
\hline
300 & $resonant$  &  $ |\cos\theta_f| < 0.92 $ 
& 1.332(2) & 1.534(2) & 1.630(2) & 1.759(2)
\\
\hline
1000 & $all$  &  $ |\cos\theta_f| < 1. $ 
& 2.619(5) & 3.040(5) & 3.278(5) & 4.132(8)
\\
\hline
1000 & $resonant$  &  $ |\cos\theta_f| < 1. $ 
& 2.588(3) & 2.978(4) & 3.167(4) & 3.576(5)
\\
\hline
1000 & $all$  &  $ |\cos\theta_f| < 0.98 $ 
& 0.802(2) & 0.923(2) & 0.981(2) & 1.065(2)
\\
\hline
1000 & $resonant$  &  $ |\cos\theta_f| < 0.98 $ 
& 0.802(2) & 0.922(2) & 0.982(2) & 1.159(2)
\\
\hline
1000 & $all$  &  $ |\cos\theta_f| < 0.92 $ 
& 0.251(1) & 0.289(1) & 0.307(1) & 0.336(1)
\\
\hline
1000 & $resonant$  &  $ |\cos\theta_f| < 0.92 $ 
& 0.2506(9) & 0.288(1) & 0.306(1) & 0.371(1)
\\
\hline
\hline
\end{tabular}
\end{center}
\caption {
Rates for $ \gamma \gamma \rightarrow \nu_e e^-
u \bar d
$ for center of mass energies of 300  and 1000 GeV.
Everything is as in table 3.
In the {\it narrow width approximation} 
the cross section at 300 and 1000 GeV is 2.99 and 3.40 $pb$ respectively.
}
\label{tggcut1}
\end{table}

Both the impact of finite W width and of non resonant
contribution are clearly seen:

$i)$ the cross section for (\ref{pro}) is about 10 \% smaller
than that for (\ref{ww})
with an invariant mass cut $\mu \le 15\div 20$ GeV
and for $20 \ {\mathrm GeV} > \mu > 10 \ {\mathrm  GeV}$  the contribution
to the cross section is still at the level of one per cent per GeV;

$ii)$ the difference among the computations which do
and do not include non-resonant contributions
\footnote
{Although the ALPHA algorithm does not make use of
the Feynman graphs technique to compute the scattering amplitudes,
it is still possible to isolate the contribution of a
subset of graphs: in the present case it can be seen, by direct
inspection,
that setting to zero the couplings of the fermions with the photon,
one indeed isolates the contribution of the doubly 
resonant diagrams only.
}
is shown in Fig.~2: already for $\mu \le 4$ GeV it is about 
1 \%, it becomes 2 \% at $\mu \le 10$  GeV
and it is larger than 10 \% for the total cross section.

The effect of angular cuts on the emitted fermions is 
also manifest.
A large fraction of the cross section occurs in correspondence
of fermions emitted along the beam direction. 
This is due to the dominant contribution of $t$ channel 
virtual W exchange
which increases the amplitude for 
the emission of `W' collinear to the photons.
The two W, in
turn, being quite strongly boosted in the laboratory frame,
emit their decay products mainly in the forward direction.

When invariant mass cuts are applied,
the non resonant contribution is almost unimportant
for $ |\cos\theta_f| < 0.98 $, $\theta_f$ being the smallest
of the angles of charged fermions with the beam direction;
this demonstrates that the bulk of the contribution of non resonant
diagrams occurs for small $\theta_f$, when an 
almost on shell virtual fermion is exchanged in the $t$
channel.

The difference is below 1 \% when angular
and invariant mass
cuts are applied ($\mu<15\div 20$ GeV, $|\cos\theta_f|<0.98$) and becomes
a few per cent when no invariant
mass cuts are applied (this can be relevant
for purely leptonic events since the W invariant mass
cannot be reconstructed in this case) 
perhaps suggesting
that for large W virtuality some unitaryty violation, induced by the
lack of gauge invariance of the subset of 
resonant diagrams, might play a role.

In Fig.~3 the angular distribution for the $u$ quark is plotted
with various choices of angular and invariant mass cuts. There
is  a small  difference
 among the `full' (including the contribution of
all graphs) and the `resonant' (including only the 
contribution of doubly resonant graphs) calculation.
When angular and invariant mass
cuts are applied 
 the difference
is likely to be statistically meaningless.
In Fig.~4 the relative  difference among the 
two distribution is reported.

To provide a semiquantitative assessment of the difference among
the two spectra I have divided the 
variable $\cos \theta_u$ ($\theta_u$ is the angle of the u 
quarks with the beam direction)
in $k$ (unequal) bins and defined the following function
\be
f = \frac {1} {N^{res}} \sum_{j=1}^k \frac { ( n_j^{all} - n_j^{res} )^2 }
{ n_j^{res} }
\label{chi}
\ee
where $n_j^{all} $ and $ n_j^{res} $ are the numbers of
events in the $j$-$th$ bin predicted according to the 
full and resonant calculation respectively and
$N^{res} $ is the total number of events according
to the resonant calculation.
Under the assumption that $n_j^{all} $ is the prediction
of a `model' and $n_j^{res} $ are `experimental'
measurements, the quantity $N^{res} f$ would be 
\footnote{This is true for large $n_j^{res}$
since in this case the error of the experimental measurement
can be estimated as $\sqrt {n_j^{res}}$. Moreover 
the expected 
spread in the experimental measurements is not accounted for
in eq. (\ref{chi}). In view of the modest purpose of
the discussion all these details are inessential.
}
the $\chi^2$
of the experimental measurements 
versus the `model' and therefore we 
can interpret $f$ as a measure of how well
the resonant calculations fits the full one. In practice
one has to keep in mind that both $n_j^{all}$ and $n_j^{res} $
(see table \ref{tggcut})
have errors and therefore $f$ is intended
to give an idea rather than an accurate statement
about the possibility to discriminate among the two spectra.
In particular since the calculation has
an error in the range $1\div2 \cdot 10^{-3}$
$f$ is not meaningless only when
one assumes a number of events smaller than $10^{6}$.
The values of $f$ as well as the `$2\sigma$' limit for $N^{res}f$
are given in table \ref{tchi}.

\begin{table}
\begin{center}
\begin{tabular}{|c|c|c|c|c|c|c|}
\hline\hline
angular cut ($e,d$)
& angular cut ($u$) & $N_{bin}$ &
$f_{\mu\le 5} $ & $f_{\mu\le 10} $ & $f_{\mu\le 20} $
& $\chi^2$ limit 
 \\
\hline\hline
$\cos\tilde\theta_f< 0.90$  & $\cos\theta_u< 0.98$ 
& 34 &
$1.67 \ 10^{-4} $ & $ 1.33 \ 10^{-4} $
& $1.26 \ 10^{-4} $ & $ N f < 58$
\\
\hline
$\cos\tilde\theta_f< 0.98$  & $\cos\theta_u< 0.98$ 
& 34 &
$9.09 \ 10^{-5} $ & $ 7.62 \ 10^{-5} $
& $7.55 \ 10^{-5} $ & $ N f < 58$
\\
\hline
$\cos\tilde\theta_f< 0.90$  & $\cos\theta_u< 0.91$ 
& 24 &
$6.96 \ 10^{-5} $ & $ 5.66 \ 10^{-5} $
& $5.49 \ 10^{-5} $ & $ N f < 44.5$
\\
\hline
$\cos\tilde\theta_f< 0.98$  & $\cos\theta_u< 0.91$ 
& 24 &
$8.83 \ 10^{-5} $ & $ 8.02 \ 10^{-5} $
& $8.07 \ 10^{-5} $ & $ N f < 44.5$
\\
\hline \hline
\end{tabular}
\end{center}
\caption {
The function $f$ defined in (4)is given
as a function of various invariant mass and angular cuts.
The angle $\tilde\theta_f$ is the minimal angle 
(in absolute value)  of $e^-$ or $\bar d$
with the
beam
direction, the angle $\theta_u$ is
the angle  of the 
$u$ quark
with respect to 
the beam direction, $f_{\mu<s}$ is 
the function $f$ defined in (4)
for a data sample to which the cut $\mu \le s$ has been applied,
$N_{bin}$ is the number of bins in which the data sample
has been divided,
finally the reported limit value for $Nf$
corresponds to a probability smaller than $0.005$
that an experimental distribution is well reproduced
by the resonant calculation. 
See the text for the caveats
of this statement.
N is the number of events and the definition of $\mu$ is given
in the caption of table 3.
}
\label{tchi}
\end{table}

The values of the same function $f$ as in (\ref{chi}),
where now the variable $\cos\theta_e$ ($\theta_e$ is
the angle of the $e^-$ with  the beam direction) is used,
are given in table \ref{tchie}.
In this case values are given also without angular cuts
for the emitted quarks, accounting for the possibility
that also partons emitted along the beam direction
might origin observable
jets
and that it might not
be possible to
reconstruct with high accuracy
the jet three momentum. When no angular cut is applied to the final
quarks the difference among the full and the resonant
calculation
is likely to be statistically meaningful unless
a tight invariant mass cut is applied.

\begin{table}
\begin{center}
\begin{tabular}{|c|c|c|c|c|c|c|}
\hline\hline
angular cut ($u,d$)
& angular cut ($e$) & $N_{bin}$ &
$f_{\mu\le 5} $ & $f_{\mu\le 10} $ & $f_{\mu\le 20} $
& $\chi^2$ limit 
 \\
\hline\hline
$\cos\tilde\theta_f< 0.90$  & $\cos\theta_e < 0.98$ 
& 34 &
$1.52 \ 10^{-4} $ & $ 1.40 \ 10^{-4} $
& $1.46 \ 10^{-4} $ & $ N f < 58$
\\
\hline
$\cos\tilde\theta_f< 0.98$  & $\cos\theta_e < 0.98$ 
& 34 &
$8.67 \ 10^{-5} $ & $ 5.95 \ 10^{-5} $
& $5.97 \ 10^{-5} $ & $ N f < 58$
\\
\hline
$\cos\tilde\theta_f< 1.$  & $\cos\theta_e< 0.98$ 
& 34 &
$8.60 \ 10^{-5} $ & $ 2.17 \ 10^{-4} $
& $5.71 \ 10^{-4} $ & $ N f < 58$
\\
\hline
$\cos\tilde\theta_f< 0.90$  & $\cos\theta_e< 0.91$ 
& 24 &
$1.24 \ 10^{-4} $ & $ 1.22 \ 10^{-4} $
& $1.27 \ 10^{-4} $ & $ N f < 44.5$
\\
\hline
$\cos\tilde\theta_f< 0.98$  & $\cos\theta_e< 0.91$ 
& 24 &
$7.48 \ 10^{-5} $ & $ 5.66 \ 10^{-5} $
& $5.83 \ 10^{-5} $ & $ N f < 44.5$
\\
\hline
$\cos\tilde\theta_f< 1. $  & $\cos\theta_e< 0.91$ 
& 24 &
$1.02 \ 10^{-4} $ & $ 2.40 \ 10^{-4} $
& $4.95 \ 10^{-4} $ & $ N f < 44.5$
\\
\hline \hline
\end{tabular}
\end{center}
\caption {
All the definition are
the same as in table
5,
but the angle $\tilde\theta_f$ is now the minimal angle 
(in absolute value) 
of $u$ or $\bar d$
quarks
with the
beam
direction, the angle $\theta_e$ is
the angle  of the
$e^-$ with respect to the beam direction.
}
\label{tchie}
\end{table}

Another observable of  interest is the angle 
$\theta_W$ among 
the reconstructed `$W^+$'
and the beam direction. In Fig.~5 it is reported the
differential cross section as a function of $\cos\theta_W$.
The relative difference among the full and the resonant contribution
is plotted in Fig.~6. 
Again the difference is concentrated at very small
angles and at large invariant mass cuts and it seems barely
observable with a statistic of $10^{6}$ events.
In table \ref{tchiw} I  give
the values of $f$ as in (\ref{chi})
where now the bins are defined as intervals in the variable
$\cos\theta_W$.

Obviously a rigorous assessment of 
the possibility to discriminate among the two predictions
would require a  more accurate strategy
including a simultaneous fit to the whole set 
of observables and 
a realistic simulation of the experimental
condition (detectors acceptance and resolution,
$E_\gamma$ spectra, etc\dots),
however it seems that one can conclude that,
at least at the few per mille level
and for the observables considered
here, the resonant calculation
appears a good approximation of
the full one. 
This good agreement is doomed to disappear
when purely leptonic final states are considered
and the observable final states do not allow to
reconstruct the `W' invariant mass.

\begin{table}
\begin{center}
\begin{tabular}{|c|c|c|c|c|c|c|}
\hline\hline
angular cut ($e,u,d$)
& angular cut ($\theta_W$) & $N_{bin}$ &
$f_{\mu\le 5} $ & $f_{\mu\le 10} $ & $f_{\mu\le 20} $
& $\chi^2$ limit 
 \\
\hline\hline
$\cos\theta_f< 0.90$  & $\cos\theta_W< 0.98$ 
& 34 &
$1.01 \ 10^{-4} $ & $ 8.61 \ 10^{-5} $
& $1.73 \ 10^{-4} $ & $ N f < 58$
\\
\hline
$\cos\theta_f< 0.98$  & $\cos\theta_W< 0.98$ 
& 34 &
$6.31 \ 10^{-5} $ & $ 6.05 \ 10^{-5} $
& $6.03 \ 10^{-5} $ & $ N f < 58$
\\
\hline 
$\cos\theta_f< 0.90$  & $\cos\theta_W< 0.91$ 
& 24 &
$6.93 \ 10^{-5} $ & $ 5.44 \ 10^{-5} $
& $5.81 \ 10^{-5} $ & $ N f < 44.5$
\\
\hline
$\cos\theta_f< 0.98$  & $\cos\theta_W< 0.91$ 
& 24 &
$4.59 \ 10^{-5} $ & $ 3.95 \ 10^{-5} $
& $3.26 \ 10^{-5} $ & $ N f < 44.5$
\\
\hline
\hline
\end{tabular}
\end{center}
\caption {
All 
the quantities in the table
are defined as in table
5 but
the angle $\theta_f$ is the minimal angle
(in absolute value) of the charged fermions 
with the
beam
and
$\theta_W$ is the angle of the reconstructed (from $p_u$ and $p_d$)
$W^+$ with the beam.
}
\label{tchiw}
\end{table}

\subsection{Gauge Invariance}
As it stands the calculation that I have presented
is not gauge invariant. The reason is that
the introduction of a finite and running W width
as in (\ref{rw})
explicitly breaks gauge invariance.
Several strategies have been proposed to restore gauge
invariance \cite{gainvr1,gainvr2,gaimma,fudge}.
The most satisfactory appears
to be the one inspired by field theoretical arguments,
namely the inclusion in the input lagrangian
of the one loop contribution of fermion
loops to the imaginary part of the gauge bosons self couplings
\cite{gaimma}. However, although straightforward,
this strategy is a bit cumbersome to be implemented
since it requires the introduction
of complicated form factors which considerably slow down 
the computational speed.
Therefore to study the quantitative impact of 
the breaking of the gauge invariance induced
by (\ref{rw}) I have used a trick
to restore gauge invariance
which is easier to implement and does not affect
computational time, the so called {\it fudge} \cite{fudge}
scheme,
namely one first computes the matrix element
with zero W width and therefore in a fully gauge
invariant way, and only at the very end
multiplies the result for a common factor
$\lambda_{fdg}$,
\be
\lambda_{fdg}= \frac 
{
(p_1^2 - M_W^2 )(p_2 - M_W^2) }
{
(p_1^2 - M_W^2 + i \Gamma_w p_1^2\theta(p_1^2)/M_W)
(p_2 - M_W^2 + i \Gamma_w p_2^2\theta(p_2^2)/M_W)
}
\label{lfdg}
\ee
where $p_1=p_e+p_\nu$ and $p_2=p_q+p_u  $,
$p_\nu$, $p_e$, $p_u$ and $p_d$
being the neutrino, electron, $u$ and $d$ quarks four momenta respectively.

The disadvantage of this scheme is that it grossly mistreats
the non resonant contribution as well as resonant non-resonant
interference when $p_1$ or $p_2$ are close to the W mass.
However in view of the good agreement between the resonant
and the full calculation when invariant mass cuts are applied
one can hope that this does not
induce sizable errors.

The results are displayed in
table \ref{tggcut} and in Figg. 7 and 8.
When angular cuts
are applied,
no appreciable difference with respect to the 
full calculation in the running width scheme is observed
thus suggesting that the breaking of the gauge invariance
from (\ref{rw}) does not affect the numerical
results 
\footnote{This statement is obviously valid only for
the unitary gauge which is used here. Since the result
is gauge dependent, choosing an appropriate gauge,
the running width scheme can lead to an arbitrary
value for the cross section.}
in an important way.
If no cut is applied a
small difference appears
and it is not possible to determine, at this level,
whether this is due to the already mentioned caveats
of the {\it fudge} scheme or to 
the gauge violation effect. 
 Ultimately 
one has to perform the calculation
in the most appropriate way using the best gauge 
restoration scheme which is available, however
the shown comparison suggests that,
to study the experimental data,
 it is possible
to use a computation with the explicit breaking of
the gauge invariance (\ref{rw}) and only
at the very end of the analysis to
check the results
with a theoretically more satisfactory calculation.

\section {Conclusions and Remarks}
I have addressed the calculation of
the process $\gamma\gamma\rightarrow e^- \bar \nu_e u \bar d $
at the tree-level by means of 
the ALPHA code.

The rate is dominated by W pairs production
and decay and a sizable contribution
comes also from the exchange
of an almost on shell $t$ channel virtual charged fermions
(always linked to the emission of a charged fermions
collinear to the incoming photons).
At the high center of mass
energies ($0.5\div 2$ TeV) typical
of future linear colliders the W bosons 
are boosted in the laboratory frame
and, since they are produced mainly in the beam direction
because of the dominant contribution from $t$ channel virtual
W exchange, also their decay products are
emitted mainly along the beam direction.

The result of the full calculations exhibits important
differencies with respect to the {\it narrow width}
approximation (see (\ref{ww}) ) which has been used
up to now in the discussion in the literature.
There is an important dependence on the cuts on the invariant
mass of the reconstructed W: requiring this mass to be
within a 5, 10 and 18 GeV interval of the measured W mass
one 
obtains differencies of about 16 \% and 24 \% for
the total rate.
Also the approximation based on the subset of
doubly resonant diagrams appears in general to be unsatisfactory
leading to discrepancies larger than 10 \%.
However when {\it both} invariant mass and angular cuts are
applied (this is impossible for purely leptonic final states)
the complete calculation
and the the contribution of 
doubly resonant diagrams only
are much closer to each other.

A final remark is in order here:
although the computation has been presented
for a fixed center of mass energy it
is a straightforward matter to allow for a variable energy
for the colliding photons and therefore 
the whole calculation
can easily be adapted to become
an event generator for the future $\gamma\gamma$
collider.

\vskip 40pt
\noindent {\Large \bf   Acknowledgments}\\
\vskip 9pt
%\acknowledgments
I thank the {\it Associazione per lo Sviluppo
della Fisica Subatomica,} {\it Ferrara} for financial support
and {\it INFN,} {\it sezione di Ferrara} for making available
computing facilities.

\newpage
\noindent\hskip -7pt
{\Large {\bf Figure Captions}}

\vskip 20pt
\begin{list}{P}{\labelwidth=100pt} 

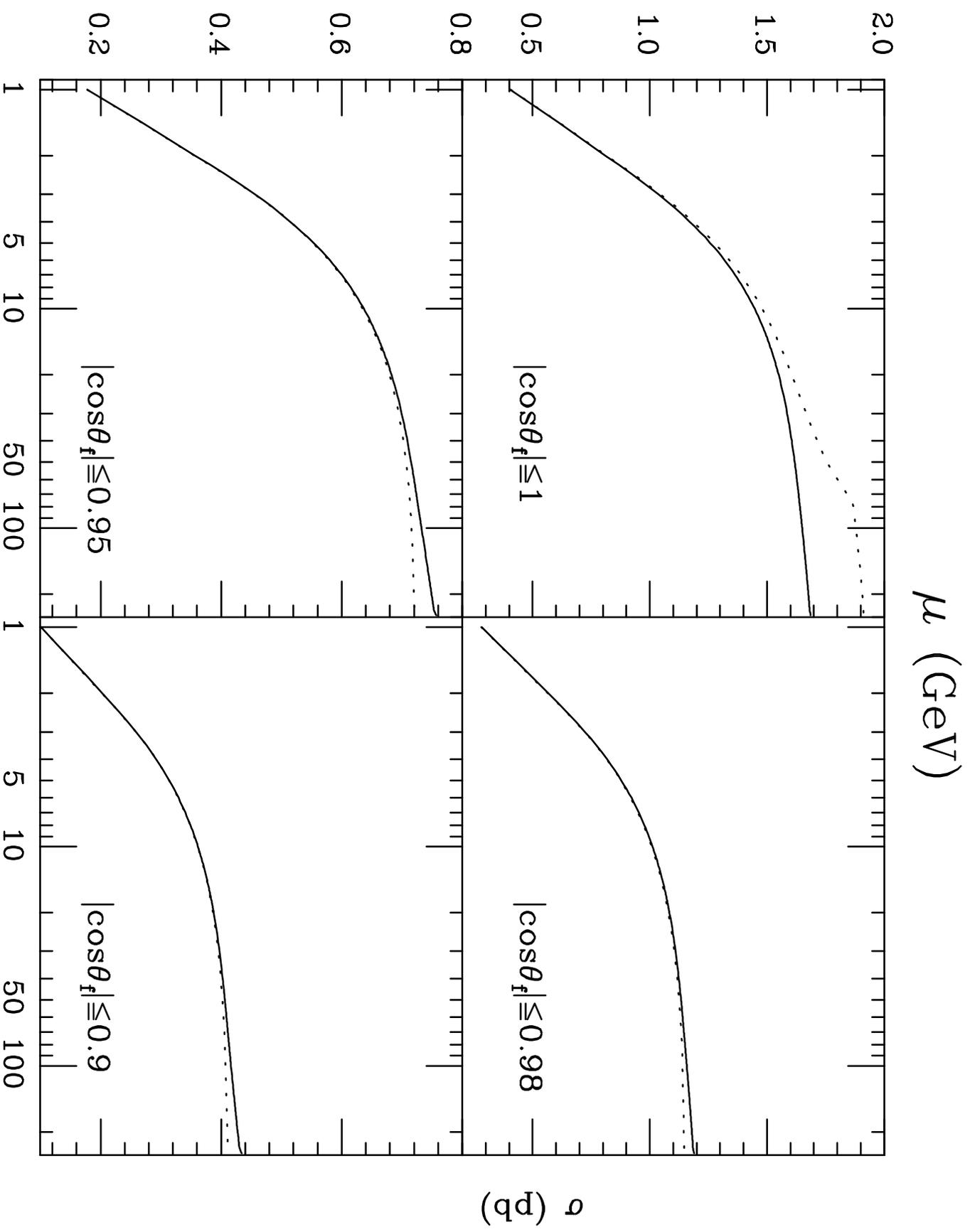
\item [Fig.~1] Differential cross section in picobarns as
a function of the invariant mass cut $\mu$ for various
angular cuts. $\mu$ and $\theta_f$ are defined as in table
\ref{tggcut}. The continuos line is the contribution
of doubly resonant diagrams only (referred as $resonant$ in 
the text) and the dotted line the full calculation
(referred as $full$ in
the text).

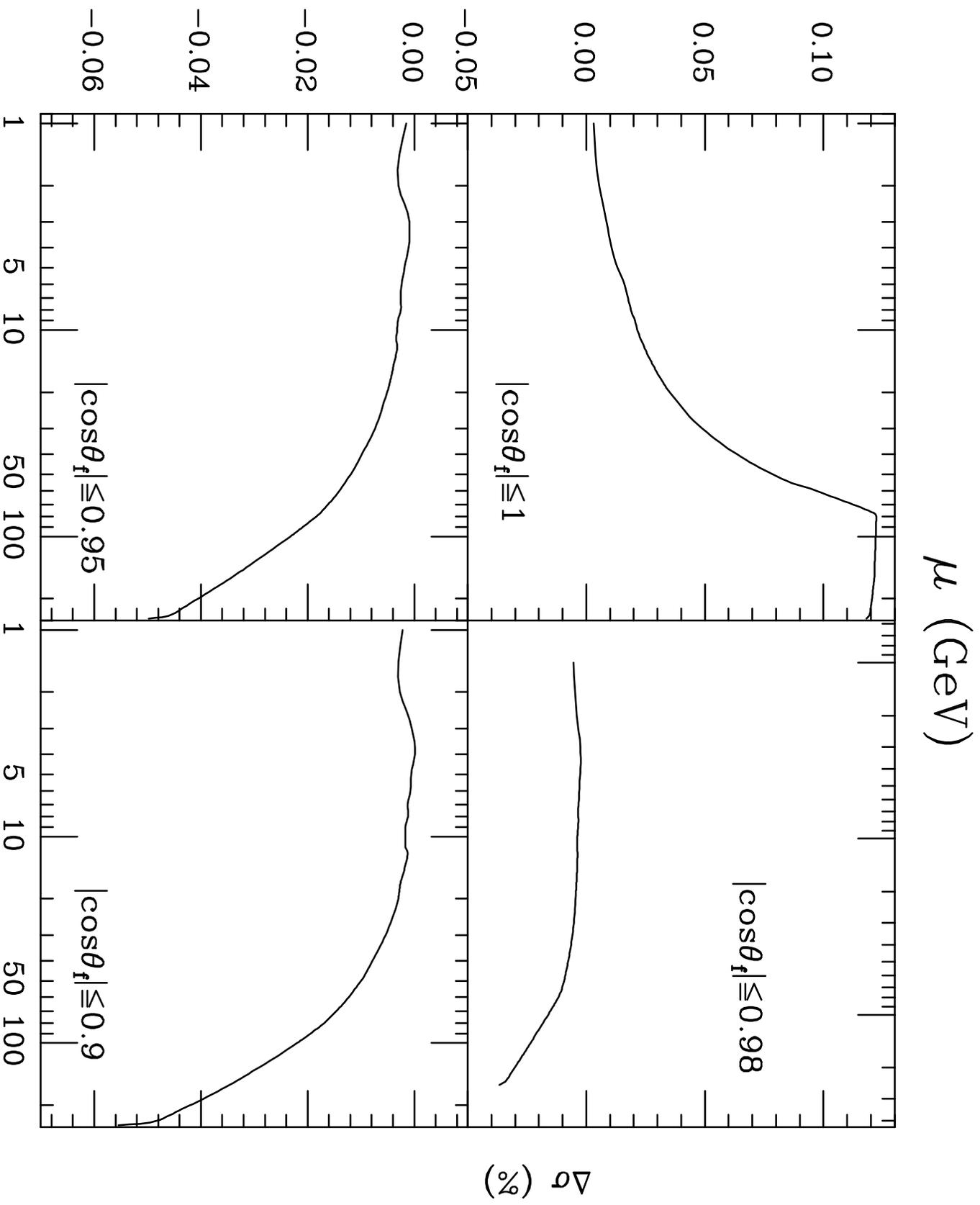
\item [Fig.~2] The relative difference among the $full$ and
$resonant$ computation (see text for the definition) as a 
function of the the invariant mass cut $\mu$ for various
angular cuts. $\mu$ and $\theta_f$ are defined as in table
\ref{tggcut},
$\Delta \sigma = (\sigma_{all} - \sigma_{resonant})/\sigma_{all} $
where the labels $all$ and $resonant$ refers to the full and resonant
computation respectively.

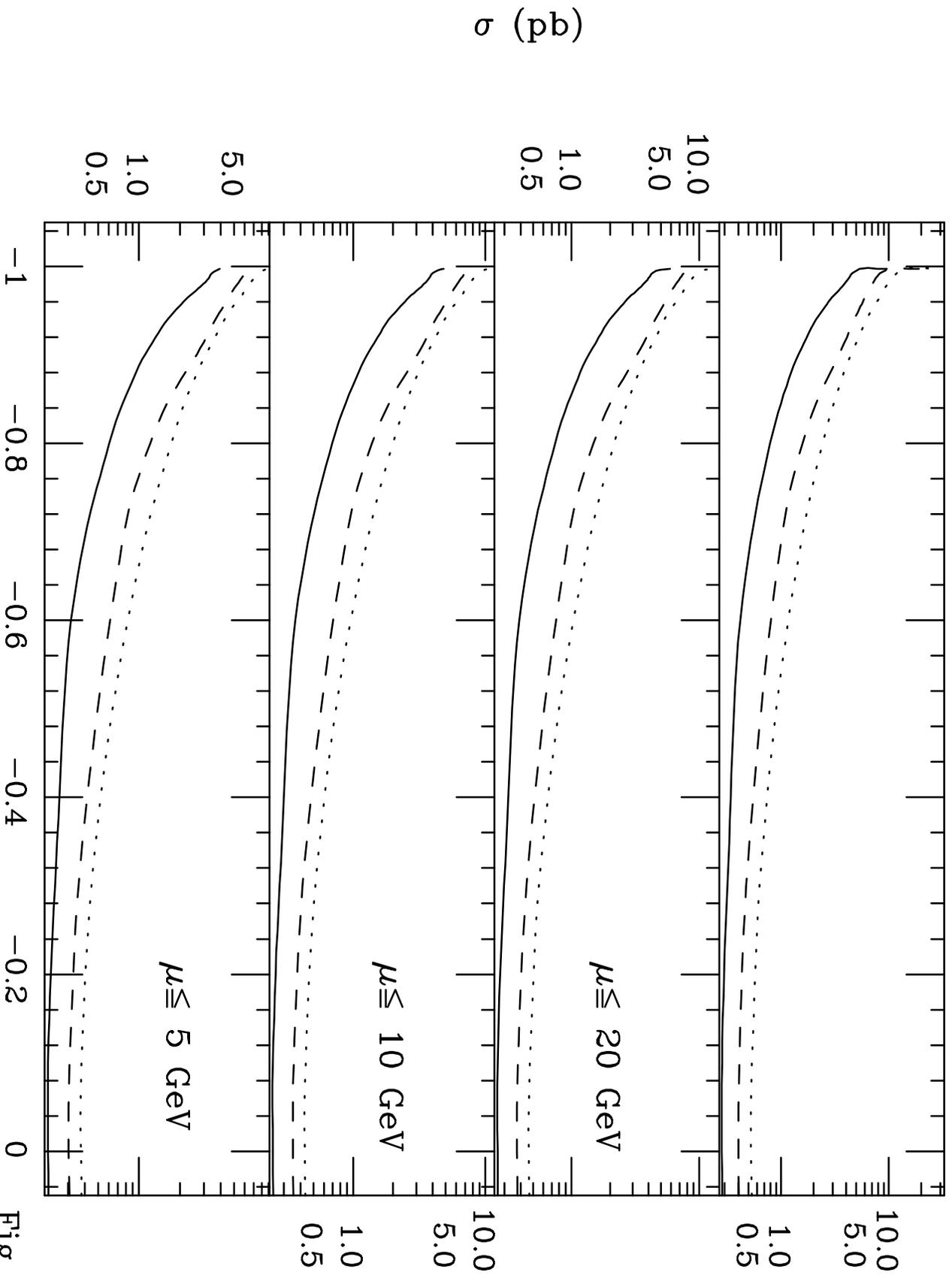
\item [Fig.~3] Differential cross section in picobarns as
a function of $\cos\theta_u$ where $\theta_u$
is the angle of the $u$ quark with respect to one of the incoming photons. 
The distribution is given for various invariant mass cuts $\mu$.
$\mu$ is defined as in table
\ref{tggcut}.
The dotted line refers to an angular cut $|\cos\tilde\theta|\le 1$
the dashed one to $|\cos\tilde\theta|\le 0.98$
and the continuos one to $|\cos\tilde\theta|\le 0.9$,
where $\tilde\theta$ is the minimal (in absolute value)
among the angles of
$e^-$ and $\bar d$ quark with the beam direction.

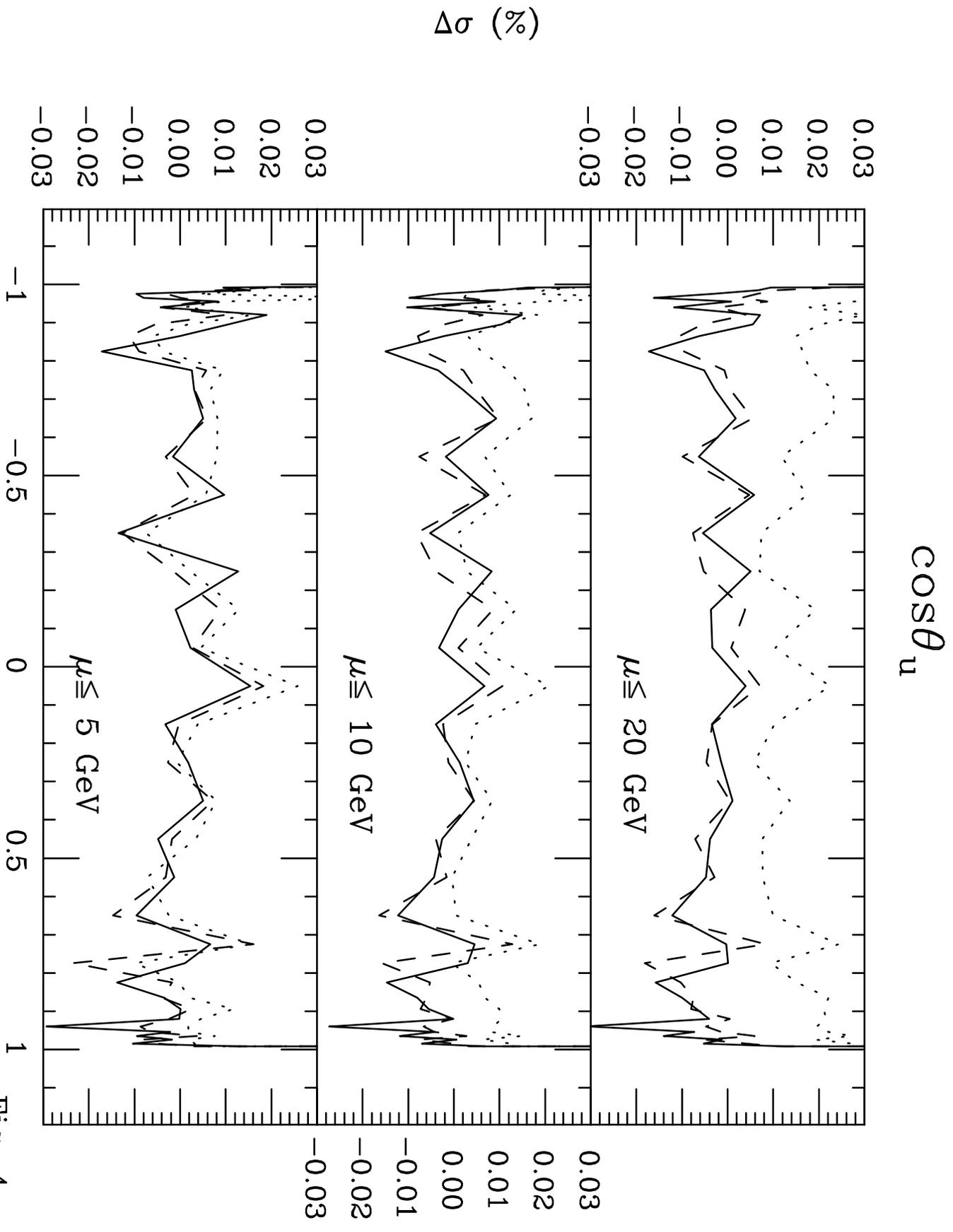
\item [Fig.~4] Relative difference among the full and resonant computation
for the differential distribution given in fig.~3.
All the quantities are defined as in fig.~3 and 
$\Delta\sigma$ as in fig.~2.

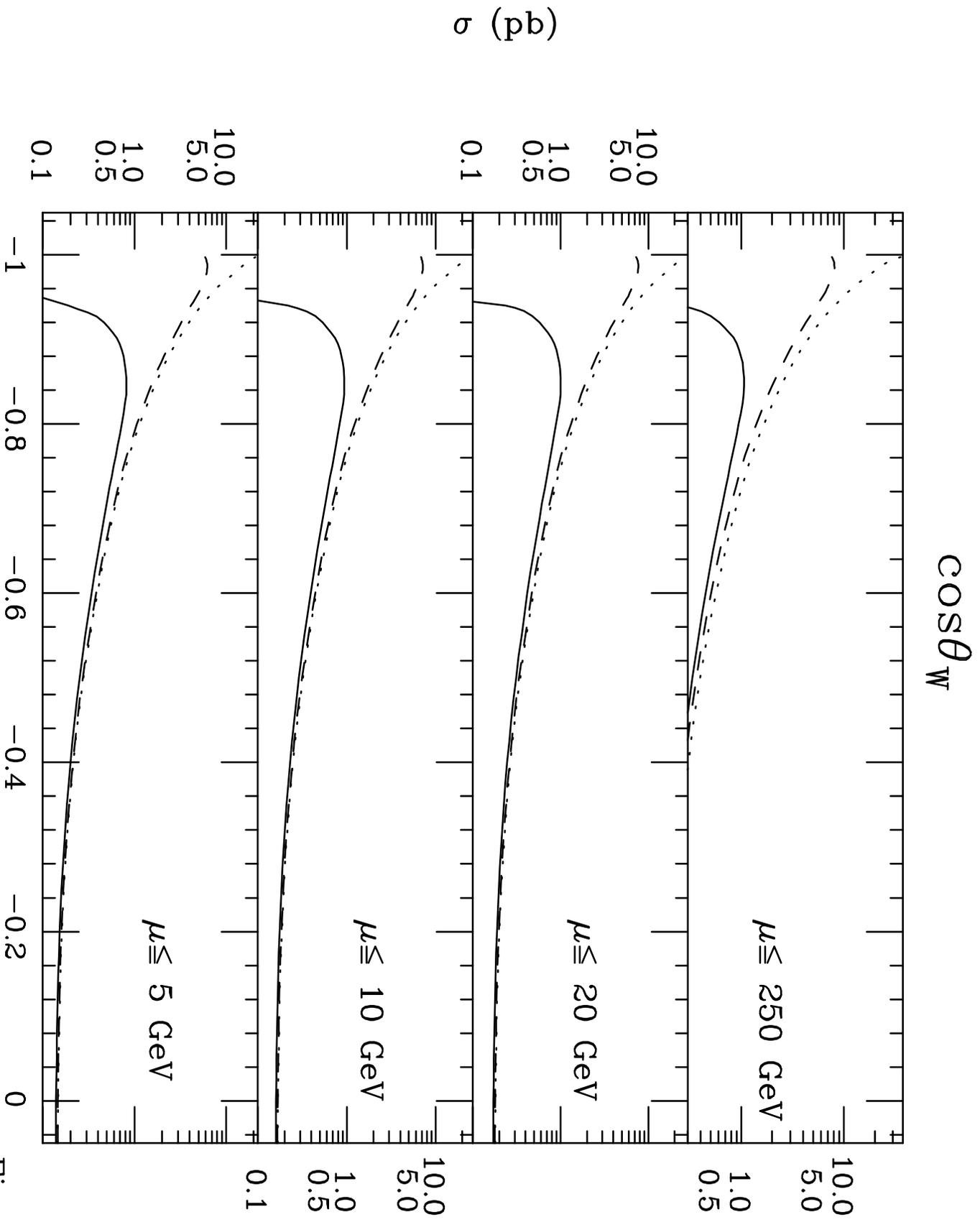
\item [Fig.~5] Differential cross section in picobarns as
a function of $\cos\theta_W$ where $\theta_W$
is the angle of the reconstructed 
W (from $u$ and $\bar d$
four momenta)
with respect to one of the incoming photons. 
The distribution is given for various invariant mass cuts $\mu$.
$\mu$ is defined as in table
\ref{tggcut}.
The dotted line refers to an angular cut $|\cos\theta_f|\le 1$
the dashed one to $|\cos\theta_f|\le 0.98$
and the continuos one to $|\cos\theta_f|\le 0.9$,
where $\theta_f$ is the minimal (in absolute value)
among the angles of
charged fermions with the beam direction.

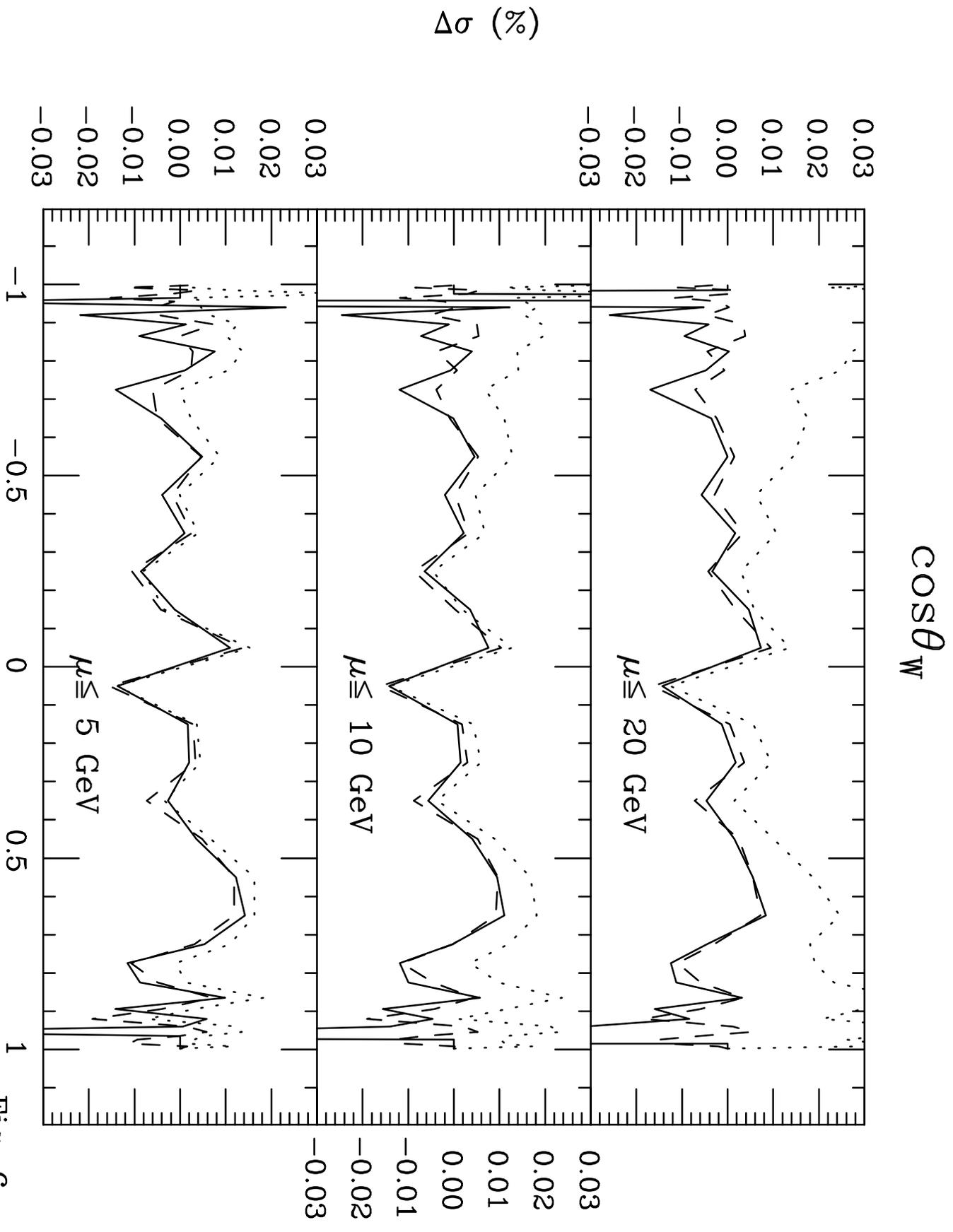
\item [Fig.~6] Relative difference among the full and resonant computation
for the differential distribution given in fig.~5.
All the quantities are defined as in fig.~5 and 
$\Delta\sigma$ as in fig.~2.

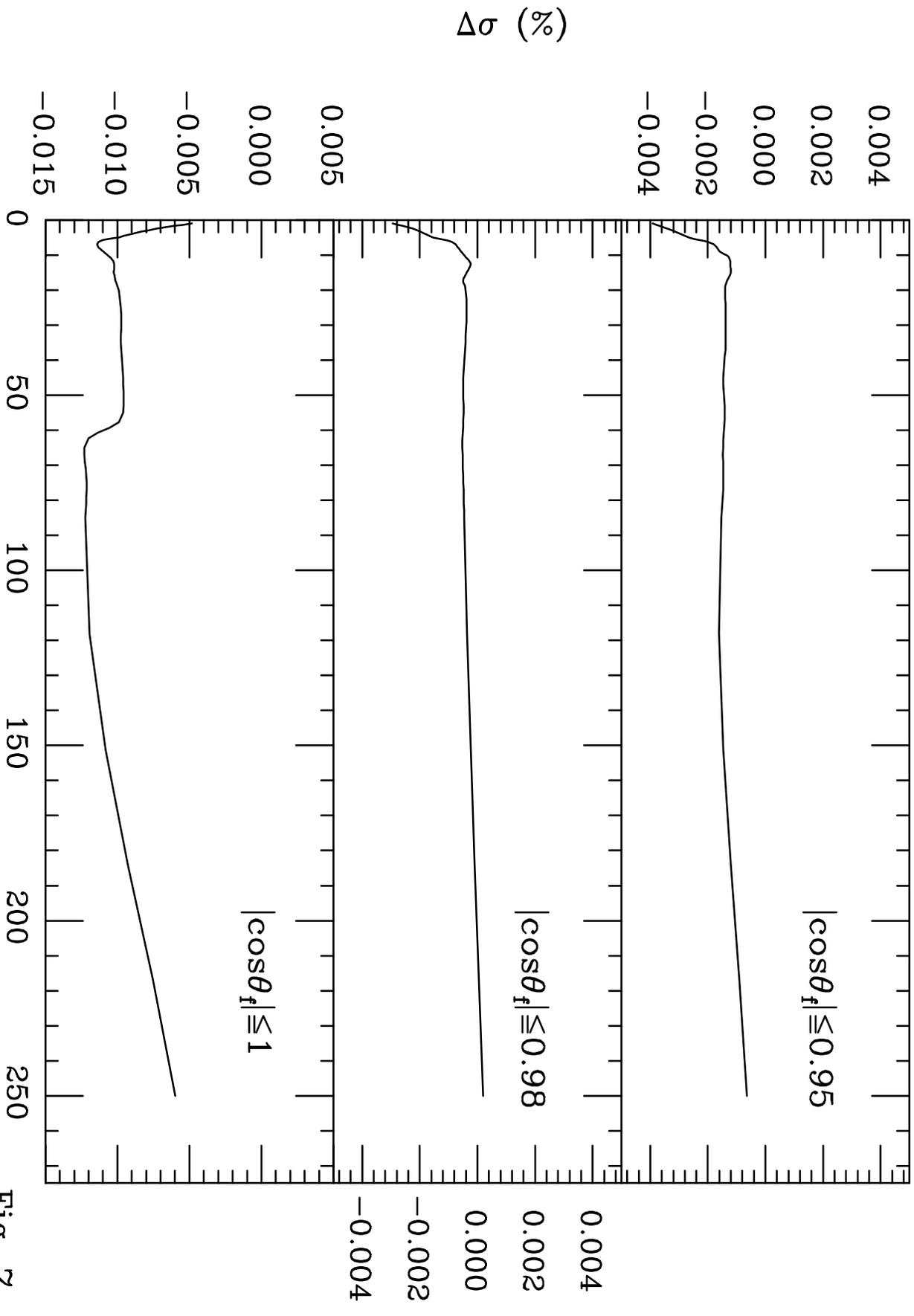
\item [Fig.~7] Relative difference among the full 
calculation with the running width for the W
propagator as in eq.~(\ref{rw})
and the calculation in the $fudge$ scheme
for the differential distribution given in fig.~2.
$\mu$ and $\theta_f$ are defined as in table \ref{tggcut}.

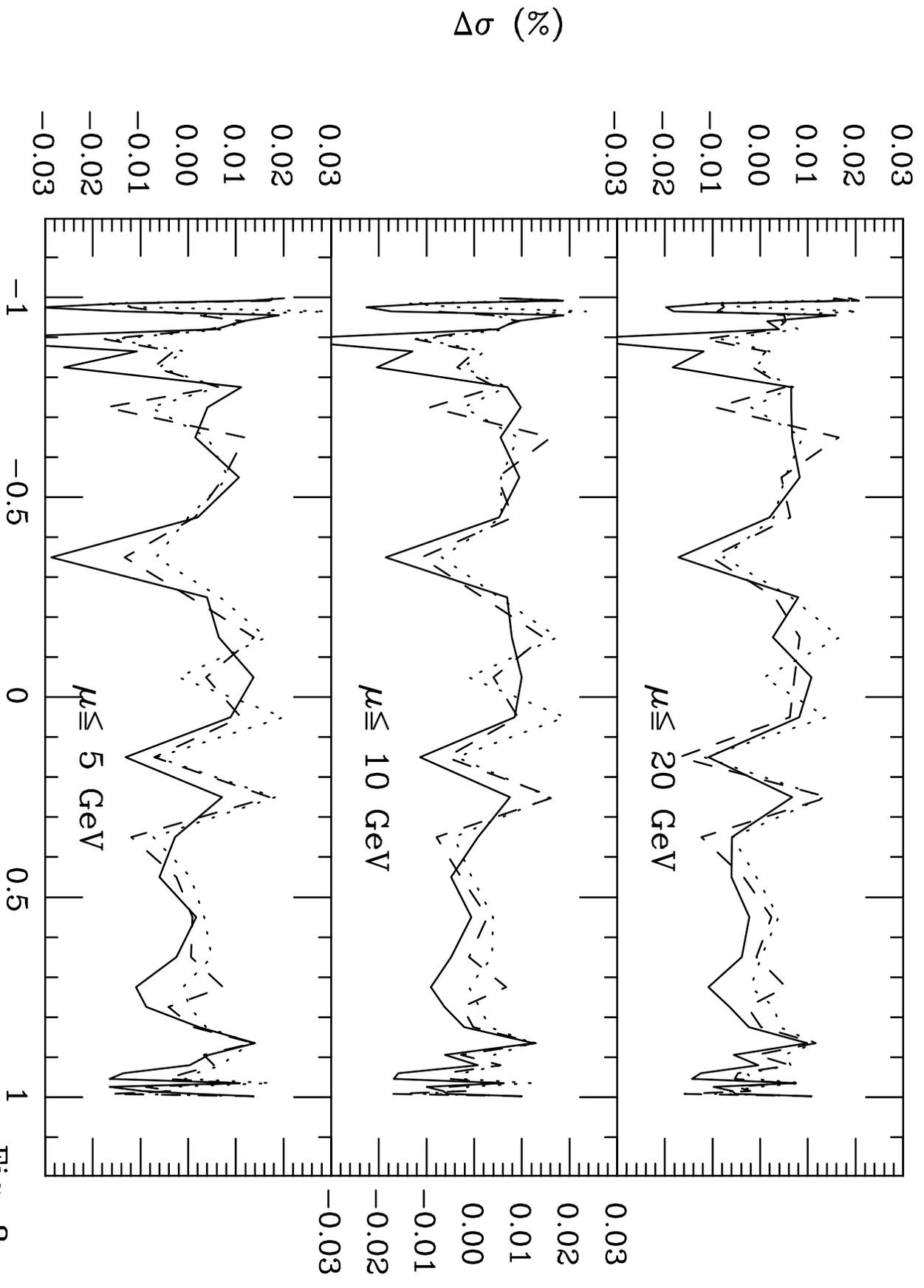
\item [Fig.~8] Relative difference among the full 
calculation with the running width for the W
propagator as in eq.~(\ref{rw}) 
and the calculation in the $fudge$ scheme
for the differential distribution given in fig.~3.
$\mu$ and $\theta_f$ are defined as in table \ref{tggcut}.

\end{list}

\end{document}